\documentclass[DIV12,11pt]{scrartcl}

\usepackage{amsthm,amssymb,amsmath}
\usepackage[noblocks]{authblk}
\usepackage[ruled, vlined, linesnumbered]{algorithm2e}
\usepackage[pdftex]{graphicx}
\graphicspath{.}
\DeclareGraphicsExtensions{.pdf,.jpeg,.png}

\usepackage{floatrow}
\newfloatcommand{capbtabbox}{table}[][\FBwidth]

\setkomafont{title}{\normalfont \LARGE \bfseries}
\setkomafont{section}{\normalfont \Large \bfseries \boldmath}
\setkomafont{subsection}{\normalfont \large \bfseries}
\setkomafont{subsubsection}{\normalfont \normalsize \bfseries}
\setkomafont{descriptionlabel}{\itshape}
\setkomafont{paragraph}{\normalfont \bfseries \boldmath}

\newtheorem{theorem}{Theorem}
\newtheorem{lemma}[theorem]{Lemma}

\title{Heuristic algorithms for obtaining Polynomial Threshold Functions with low densities}

\author{Can Eren Sezener}
\author{Erhan Oztop}

\affil{Ozyegin University, Istanbul, Turkey}

\date{\vspace*{-2em}}

\usepackage{caption} 
\begin{document}

\maketitle

\thispagestyle{empty}

\begin{abstract}

In this paper we present several heuristic algorithms, including a Genetic Algorithm (GA), for obtaining polynomial threshold function (PTF) representations of Boolean functions (BFs) with small number of monomials. We compare these among each other and against the algorithm of Oztop \cite{Oztop2009} via  computational experiments. The results indicate that our  heuristic algorithms find more parsimonious representations compared to the those of non-heuristic and GA-based algorithms.

\end{abstract}

\section{Introduction}

An $n$-variable Boolean function $f: \{-1, 1\}^n \rightarrow \{-1, 1\}$ is said to be sign-represented by $p(\mathbf{x})$ if $sign(p(\mathbf{x}))=f(\mathbf{x})$ for all $\mathbf{x} \in \{-1, 1\}^n$. In this case, we also say that BF $f$ is realized by a PTF. For a given BF $f$, there are infinitely many sign-representing polynomials which can have different number of monomials. A PTF for a given BF can be obtained by various algorithms; however, no known algorithms can guarantee a PTF with the minimum number of monomials without conducting an exhaustive search. Oztop gave the first algorithm that guarantees a PTF with $0.75\times2^n$ monomials or less \cite{Oztop2009}.  Later, Amano showed that almost all BFs can be sign-represented with less than $0.617\times2^n$ monomials  \cite{Amano2010}. Recently, Sezener and Oztop proposed a heuristic algorithm that obtains very efficient sign-representations by taking advantage of spectral coefficients \cite{Sezener2015}. In this study, we not only investigate the performance of this algorithm but also propose a faster version of it, albeit with some increase in the number of monomials found. Furthermore, we also investigate how genetic algorithms would perform in searching for a parsimonious  sign-representation.  Finally, these three methods, are compared with each other and with the non-heuristic algorithm of \cite{Oztop2009}.

\section{Algorithms for sign-representation}

In this section, we give basic definitions and the algorithms for obtaining PTFs.


\subsection{Representation of Boolean Functions:}
For every BF $f$, it can be shown that there is a unique multilinear polynomial $p_f$ that exactly represents $f$:
$p_f({x_1},{x_2}, \cdots ,{x_n}) = \sum\limits_{i = 1}^{{2^n}} {{s_i}\prod\limits_{k \in {S_i}} {{x_k}} }$ 
where $S_i$ runs over the powerset of \{1,2,..,n\}. This representation is called the spectral representation of $f$, and the coefficients are called as the spectral coefficients. With a fixed ordering over the monomials (i.e. the products appearing in the expression of $p_f$), the spectrum (i.e., the collection of the spectral coefficients) can be considered as a vector ${\mathbf{s}}\in{\mathbb{R}^{{2^n}}}$, and used to represent $f$. The spectrum can be obtained by Lagrange Interpolation \cite{Oztop2006}. Note that the with this, we obtain two vector representations of a given BF: one is the spectrum ${\mathbf{s}}$ and the other is the natural binary vector ${\mathbf{f}}$ that is obtained by listing the function values for all variable assignments. Now, with the adoption of a suitable ordering, this two representations can be related nicely: ${\mathbf{f}} = {{\mathbf{D}}^{n}}{\mathbf{s}}$ where $\mathbf{D}^{n}$  is the $2^n\times2^n$ Sylvester-type Hadamard Matrix with columns representing the monomials evaluated at all possible input assignments taken in the adopted assignment  order. 



Polynomial representation of Boolean functions naturally extends to \textit{sign-representation}: instead of requiring exact interpolation we ask only the sign of the polynomial to agree with the function at each input combination. We say a polynomial $p$ sign-represents a Boolean function $f$, if and only if $f({x_1},{x_2}, \cdots ,{x_n}) = {\mathop{\rm sgn}} (p({x_1},{x_2}, \cdots ,{x_n}))$ for all ${x_i} \in \{-1,1\}$. In vector notation this is equivalent to saying ${\mathbf{f}} = {\mathop{\rm sgn}} ({{\mathbf{D}}_n}{\mathbf{a}})$ where ${\mathbf{a}}$ is the coefficients of $p$ ordered as in the spectrum definition. It can be shown that for a given Boolean function  $f$, all the solutions (the coefficients of sign-representing polynomials) are of the form ${\mathbf{a}} = {2^{ - n}}{{\mathbf{D}}_n}{{\mathbf{Y}}_f}{\mathbf{k}}$ with arbitrary ${\mathbf{k}} > {\mathbf{0}}$. 
From this one can obtain the following useful result (\cite{Sezener2015} and \cite{Oztop2009}):

\begin{lemma}
Let ${{\mathbf{Q}}_f} = diag({\mathbf{f}}){{\mathbf{D}}_n} $, and $\mathbf{A}$, $\mathbf{B}$ be matrices made up from an arbitrary partition of the columns of ${{\mathbf{Q}}_f}$. Then  $\exists \mathbf{k>0}$ such that $\mathbf{B}^T\mathbf{k = 0}$ if and only if  
$\mathbf{a}=[\mathbf{A 0}]^T\mathbf{r}$ with some $\mathbf{r>0}$ is a solution for ${{\mathbf{Q}}_f}{\mathbf a}>{\mathbf 0}$. That is, if we can find a $\mathbf{k>0}$ such that  $\mathbf{B}^T\mathbf{k = 0}$, it means that we can eliminate the monomials which correspond to the rows in $\mathbf{B}$.
\end{lemma}
The minimum number of monomials that would be sufficient to represent a given BF $f$ is called the \textit{threshold density} of $f$. A brute force algorithm to find this value is easy to give: 

\begin{table}[h]
\begin{algorithm}[H]
Enumerate all the column submatrices of $\mathbf{Q}$ as $\mathbf{Q}_1, \mathbf{Q}_2,...\mathbf{Q}_{2^{2^n}}$\;
Set $S$ as $\{\}$\;
 \For{$i = 1..2^{2^n}$}{
  \If{$\mathbf{YQ}_i\mathbf{a} > \mathbf{0}$ is satisfiable}{
   store (\# of columns of $\mathbf{Q}_i$) in $S$\;
   }
}
\Return{min(S)}
\caption{The brute-force algorithm for a n-variable BF }
\end{algorithm}
\end{table}


Although, the satisfiability check in Line 4 can be done efficiently (e.g. by using Linear Programming methods), due to the super-exponential growth (with respect to $n$) of the number of column submatrices, the brute-force solution is not feasible for investigating threshold densities of BFs with $4$ or more variables. It is worth noting that as each column of $\mathbf{Q}_i$ corresponds to a monomial, Lemma 1 tells us that the satisfiability of $\mathbf{YQ}_i\mathbf{a} > \mathbf{0}$   can also be shown by finding a positive vector that is orthogonal to the monomials not appearing in  $\mathbf{Q}_i$. In the following section this property is used and instead of talking about satisfiability we talk about elimination of monomials by inner product with a positive vector. 

\subsection{Heuristic algorithms for sign-representation} 
Recently, Sezener and Oztop proposed a fast heuristic algorithm to obtain sign-representations with a very few number of monomials \cite{Sezener2015}. 
The heuristic (see Algorithm 2) is based on the intuition that some monomials are `easier' to eliminate and the algorithm first attempts to eliminate the `easy' monomials and then moves on to the harder ones. It does a single pass over the $2^n$  monomials and therefore is very efficient compared to the brute-force method which searches the power set of monomials (of cardinality $2^{2^n}$) for solutions.


\begin{algorithm}[H]
 m=1, E=\{\}\;
 Sort the monomials using their spectral coefficients as the sorting key\;
 \While{$m < 2^n$}{
  \If{monomials from 1 to $m$ that are not included in $E$ can be eliminated}{
   add $m$ to E\;
   }
   $m = m+1$\;
 }
 \Return{m}
 \caption{L-Heuristic}
\end{algorithm}

We propose a modification to this algorithm which increases its speed. The idea is to change $m$ (which controls how many  monomials are included in the elimination list $E$) somewhat similar to binary search, rather than incrementing it by one at each trial ($2^n$ many increments in total). With this change, the subroutine that checks for eliminability (e.g. a LP routine) would be called $\log_2{2^n} = n$ times in the worst case as opposed to $2^n$ times. 

\begin{algorithm}[H]
 lo = 1, hi = $2^n-1$\;
 \While{$lo \le hi$}{
  $m=floor((hi+lo)/2)$\;
  \eIf{first $m$ monomials can be eliminated}{
   $lo=m+1$\;
   }{
   $hi=m-1$\;

  }
 }
 \Return{m}
 \caption{B-Heuristic}
\end{algorithm}

\subsection{Genetic Algorithms for sign-representation}
We can cast the problem of finding the threshold density of a BF $f$ as an optimization problem. Let $\mathbf{b}$ be an indicator binary vector of length $2^n$ to uniquely identify a submatrix $\mathbf{Q_b}$ of $\mathbf{Q}$ for elimination by having a $1$ at position $i$ to pick the $i^{th}$ column of $\mathbf{Q}$ ($b$ in fact represents the elimination set $E$ in Algorithms 1 and 2). Now, our goal is to find the $\mathbf{b}$ with the maximum number of 1's that indicates a $\mathbf{Q_b}$ that can be eliminated, thus ensures a solution due to Lemma 1. Genetic Algorithms (GAs) at the onset, seem to be suitable for obtaining sign-representations as there are multiple global minima, and the relation among the monomials (i.e.  relation among the columns of $\mathbf{Q}$), suggests that the cross-over operation of GA  may do a good job at reaching better solutions from the existing solutions. In the current implementation, we used a simple binary GA \cite{binary_GA}. The GA searches for genotypes (i.e.  $\mathbf{b}$ vectors) with higher fitness values, which is simply defined as $\sum_{i=1}^{2^n}b_i$ if $isEliminable(\mathbf{Q_b})$ else $0$, where ${\it isEliminable}$ checks whether the columns of its argument can be annihilated by a positive vector. 
The results of GA for sign representation  (we used 16 chromosomes with 0.01 mutation rate for 100 generations) is given in the next section, together with other algorithms. 

\section{Results}
\begin{figure}[h!]
\centering
\includegraphics[width=\textwidth]{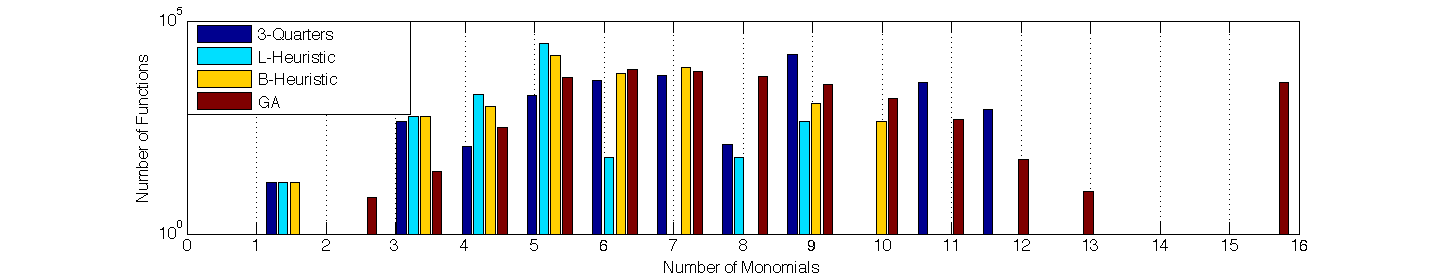}
\caption{Distribution of the number of monomials to represent 4-variable BFs (log-lin scale)}
\label{hist}
\end{figure}

We found PTFs for $4$-variable BFs using the aforementioned algorithms. Table 1 shows the average number of monomials and the average computation durations (for a standard PC).
\begin{table}[h]
\label{comp}
\small
\begin{center}
\begin{tabular}{l l l}
\hline
Algorithm & Avg. \# monomials & Avg. computation time (s) \\
\hline
3-Quarters & 8.2720 & 0.0007 \\
L-Heuristic & 4.9678 & 0.0654 \\
B-Heuristic & 5.8115 & 0.0199 \\
GA & 7.9941 & 4.2678 \\
\hline
\end{tabular}
\caption{Result comparison}
\end{center}
\end{table}%
It is also useful to see the distribution of densities. Figure \ref{hist} shows the number of monomials in PTFs obtained by the four algorithms for the first half of $4$-variable BFs. For example, it can be seen that the B-Heuristic solved approximately $10^3$ of the problems with $4$ monomials. 
As an example, Table 2 gives the solutions found by the algorithms compared for one of the hardest 4-variable BFs to sign-represent  \cite{Sezener2015}.

\begin{table}[h!]
\caption{PTF representations of $f = [-1, -1,  -1,  -1,    -1,    -1,     1,     1,    -1,     1,    -1,     1,    -1,     1,     1,    -1]$}
\small
\begin{center}
\begin{tabular}{l l}
\hline
Algorithm & Polynomial representation \\
\hline
3-Quarters & $ -2 \cdot x_0 -2 \cdot x_1 \cdot x_0 -2 \cdot x_2 \cdot x_0 +2 \cdot x_2 \cdot x_1 \cdot x_0 -3 \cdot x_3 +3 \cdot x_3 \cdot x_0 -3 \cdot x_3 \cdot x_1 $ \\
 & $+3 \cdot x_3 \cdot x_1 \cdot x_0 -3 \cdot x_3 \cdot x_2 +3 \cdot x_3 \cdot x_2 \cdot x_0 +3 \cdot x_3 \cdot x_2 \cdot x_1 -3 \cdot x_3 \cdot x_2 \cdot x_1 \cdot x_0$ \\
L-Heuristic & $ -   x_1 \cdot x_0 -   x_2 \cdot x_0 +  x_2 \cdot x_1 \cdot x_0 -   x_3 \cdot x_1 +  x_3 \cdot x_1 \cdot x_0 -   x_3 \cdot x_2$  \\
 & $+  x_3 \cdot x_2 \cdot x_0 +  x_3 \cdot x_2 \cdot x_1 -   x_3 \cdot x_2 \cdot x_1 \cdot x_0$ \\
B-Heuristic & $ -2 \cdot x_1 \cdot x_0 -   x_2 -   x_2 \cdot x_0 +2 \cdot x_2 \cdot x_1 -   x_3 +2 \cdot x_3 \cdot x_0 -   x_3 \cdot x_1 -2 \cdot x_3 \cdot x_2 $ \\
 & $+  x_3 \cdot x_2 \cdot x_0 +  x_3 \cdot x_2 \cdot x_1$ \\
GA & $ -2 \cdot x_0 -3 \cdot x_1 -2 \cdot x_2 -   x_2 \cdot x_0 -2 \cdot x_3 -2 \cdot x_3 \cdot x_1 +  x_3 \cdot x_1 \cdot x_0 -3 \cdot x_3 \cdot x_2$ \\
 & $+2 \cdot x_3 \cdot x_2 \cdot x_0 -2 \cdot x_3 \cdot x_2 \cdot x_1 \cdot x_0
$ \\
\hline
\end{tabular}
\end{center}
\label{eg}
\end{table}%


\section{Conclusion}

From the computations carried out it can be seen that the L-Heuristic finds representations with the least number of monomials among the studied four algorithms. The B-Heuristic gets very close to the L-Heuristic, which is an impressive performance considering the reduced computation time.
Therefore, the  B-Heuristic seems to be a good choice studying high dimensional BFs. It is surprising that GAs that perform very well in a wide range of search problems fail to produce good results for the minimal sign-representation problem. The 3-Quarters algorithm performs comparable to the GA in terms of number of monomials found; however, the 3-Quarters uses a fraction of the time the GA uses as it is not based on a search heuristic. It is an interesting open problem in evolutionary computation to come up with domain specific crossover and mutation parameters to improve the performance of GAs for sign-representation.

\bibliographystyle{unsrt}
\bibliography{Sign-Representation.bib}

\end{document}